\documentclass[sigconf,authordraft,review=false,authordraft=false, anonymous=false,timestamp=false]{acmart}
\usepackage{balance} 
\usepackage{color}
\usepackage{multirow}
\usepackage{relsize}
\newcommand{\amin}[1]{\textcolor{black}{#1}}

\newcommand{\arian}[1]{\textcolor{black}{#1}}

\usepackage{comment}
\usepackage{amsmath}
\usepackage{accents}
\usepackage{hyperref}
\usepackage{enumerate}

\usepackage[shortlabels]{enumitem}
\AtBeginDocument{%
 \providecommand\BibTeX{{%
  \normalfont B\kern-0.5em{\scshape i\kern-0.25em b}\kern-0.8em\TeX}}}


\copyrightyear{2022}
\acmYear{2022}
\setcopyright{rightsretained}
\acmConference[ICTIR '22]{Proceedings of the 2022 ACM SIGIR International Conference on the Theory of Information Retrieval}{July 11--12, 2022}{Madrid, Spain.}
\acmBooktitle{Proceedings of the 2022 ACM SIGIR International Conference on the Theory of Information Retrieval (ICTIR '22), July 11--12, 2022, Madrid, Spain}\acmDOI{10.1145/3539813.3545133}
\acmISBN{978-1-4503-9412-3/22/07}
\usepackage{etoolbox}
\makeatletter
\patchcmd{\maketitle}{\@copyrightpermission}{
   \begin{minipage}{0.3\columnwidth}
     \href{https://eur03.safelinks.protection.outlook.com/?url=https\%3A\%2F\%2Fcreativecommons.org\%2Flicenses\%2Fby\%2F4.0\%2F&amp;data=05\%7C01\%7Cm.a.abolghasemi\%40liacs.leidenuniv.nl\%7Cfa0786c51c1f4f9e8f2308da601f396e\%7Cca2a7f76dbd74ec091086b3d524fb7c8\%7C0\%7C0\%7C637927984592715552\%7CUnknown\%7CTWFpbGZsb3d8eyJWIjoiMC4wLjAwMDAiLCJQIjoiV2luMzIiLCJBTiI6Ik1haWwiLCJXVCI6Mn0\%3D\%7C2000\%7C\%7C\%7C&amp;sdata=Pe\%2B7JqOnaG8ZjeyLS3TfmdYxlE7Bd8r023jGy64l5G4\%3D&amp;reserved=0}
     {\includegraphics[width=0.90\textwidth]{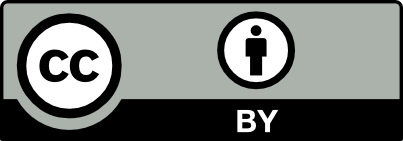}}
   \end{minipage}\hfill
   \begin{minipage}{0.7\columnwidth}
     \href{https://eur03.safelinks.protection.outlook.com/?url=https\%3A\%2F\%2Fcreativecommons.org\%2Flicenses\%2Fby\%2F4.0\%2F&amp;data=05\%7C01\%7Cm.a.abolghasemi\%40liacs.leidenuniv.nl\%7Cfa0786c51c1f4f9e8f2308da601f396e\%7Cca2a7f76dbd74ec091086b3d524fb7c8\%7C0\%7C0\%7C637927984592715552\%7CUnknown\%7CTWFpbGZsb3d8eyJWIjoiMC4wLjAwMDAiLCJQIjoiV2luMzIiLCJBTiI6Ik1haWwiLCJXVCI6Mn0\%3D\%7C2000\%7C\%7C\%7C&amp;sdata=Pe\%2B7JqOnaG8ZjeyLS3TfmdYxlE7Bd8r023jGy64l5G4\%3D&amp;reserved=0}
     {This work is licensed under a Creative Commons Attribution International 4.0 License.}
   \end{minipage}
 
   \vspace{5pt}
}{}{}

\makeatother
\begin{document}

\title{On the Interpolation of Contextualized Term-based Ranking with BM25 for Query-by-Example Retrieval}


\author{Amin Abolghasemi}
\email{m.a.abolghasemi@liacs.leidenuniv.nl}
\affiliation{%
  \institution{Leiden University}
  \city{Leiden}
  \country{Netherlands}
}

\author{Arian Askari}
\email{a.askari@liacs.leidenuniv.nl}

\affiliation{%
 \institution{Leiden University}
 \city{Leiden}
 \country{Netherlands}}

\author{Suzan Verberne}
\email{s.verberne@liacs.leidenuniv.nl}

\affiliation{%
 \institution{Leiden University}
 \city{Leiden}
 \country{Netherlands}}







\begin{abstract}
Term-based ranking with pre-trained transformer-based language models has recently gained attention as they bring the contextualization power of transformer models into the highly efficient term-based retrieval. In this work, we examine the generalizability of two of these deep contextualized term-based models in the context of query-by-example (QBE) retrieval in which a seed document acts as the query to find relevant documents. In this setting --- where queries are much longer than common keyword queries --- BERT inference at query time is problematic as it involves quadratic complexity. We investigate TILDE and TILDEv2, both of which leverage BERT tokenizer as their query encoder.
With this approach, there is no need for BERT inference at query time, and also the query can be of any length. Our extensive evaluation on the four QBE tasks of SciDocs benchmark shows that in a query-by-example retrieval setting TILDE and TILDEv2 are still less effective than a cross-encoder BERT ranker. However, we observe that BM25 could show a competitive ranking quality compared to TILDE and TILDEv2 which is in contrast to the findings about the relative performance of these three models on retrieval for short queries reported in prior work. This result raises the question about the use of contextualized term-based ranking models being beneficial in QBE setting. We follow-up on our findings by studying the score interpolation between the relevance score from TILDE (TILDEv2) and BM25.
We conclude that these two contextualized term-based ranking models capture different relevance signals than BM25 and combining the different term-based rankers results in statistically significant improvements in QBE retrieval. Our work sheds light on the challenges of retrieval settings different from the common evaluation benchmarks. It could be of value as future work to study other contextualized term-based ranking models in QBE settings.
\end{abstract}




\maketitle
\section{Introduction}
Query-by-Example (QBE) retrieval is an Information Retrieval (IR) setting in which a seed document\footnote{Throughout this paper, we use the term ``document'' to refer to a unit of retrieval \cite{lin2021pretrained}} acts as the query to represent the user's information need and the retrieval engine searches over a collection of the same type of documents \cite{sarwar2020qbe,mysore2021csfcube,mysore2021multi,abolghasemimultitask2022}. This retrieval setup is typical in professional, domain-specific tasks such as legal case law retrieval \cite{abolghasemimultitask2022,askari2021combining}, patent prior art search \cite{piroi2011clef,piroi2019multilingual,fujii2007overview}, and scientific literature search \cite{mysore2021csfcube,abolghasemimultitask2022,mysore2021multi}. While using a document as a query could become challenging due to its length and complex semantic structure, prior work has shown that traditional term-based retrieval models like BM25 \cite{robertson3m} are highly effective when used in QBE retrieval \cite{abolghasemimultitask2022,askari2021combining,rosa2021yes}.

Recently, deep contextualized term-based retrieval models have gained attention as they bring the contextualization power of the pre-trained transformer-based language models into the highly efficient term-based retrieval. Examples of such models are DeepImpact \cite{deepimpact2021}, SPLADE \cite{formal2021splade}, SPLADEv2  \cite{formal2021spladev2}, TILDE \cite{zhuang2021tilde}, TILDEv2 \cite{zhuang2021fasttilde}, COIL \cite{gao2021coil}, and uniCOIL \cite{lin2021few}. Here, we specifically investigate TILDE, which is a term independent likelihood model, and its follow-up TILDEv2 which is a deep contextualized lexical exact matching model. 

TILDE and TILDEv2, which are introduced as term-based re-ranking models, follow a recent paradigm in term-based retrieval where term importance is pre-computed with scalar term weights.
Besides, to predict the relevance score, both of these models use the BERT tokenizer as their query encoder which means that they do not need to perform any BERT inference at query time to encode the query.
However, leveraging tokenizer-based encoding of the query trades off the query representation and therefore effectiveness with higher efficiency at inference time \cite{zhuang2021fasttilde}. While the effectiveness of these models is evaluated on tasks and benchmarks where we have short queries, e.g., MSMARCO Passage Ranking \cite{nguyen2016ms} and the TREC DL Track \cite{craswell2021trec}, in this paper, we evaluate them in the aforementioned QBE retrieval setting where queries are much longer than common keyword queries. In this regard, we address the following research questions: 

\begin{enumerate}[RQ1]
\item How effective are TILDE and TILDEv2 in query-by-example retrieval?
\end{enumerate}

A specific direction in answering \textit{RQ1} is to investigate the ranking quality of TILDE and TILDEv2 in comparison with the effective cross-encoder BERT ranker \cite{nogueira2019passage,abolghasemimultitask2022}, which is described in section \ref{sec:cross-encod}. We are interested in this direction for two reasons. First, the cross-encoder BERT ranker exhibits quadratic complexity in both space and time with respect to the input length \cite{lin2021pretrained} and this is aggravated in QBE where we have long queries. TILDE and TILDEv2, however, do not need any BERT inference at query time. Second, due to the maximum input length of BERT, cross-encoder BERT ranker, which uses the concatenation of the query and the document, might not cover the whole query and document tokens in a QBE setting, whereas in TILDE and TILDEv2, the query can be of any length and documents are covered up to the maximum length of BERT.

Additionally, since TILDEv2 pre-computes the term weights only for those tokens existing in the documents, one risk is that it might aggravate the vocabulary mismatch problem. A typical approach to address this issue is to use document expansion methods. \citet{zhuang2021fasttilde} use TILDE as their document expansion model for TILDEv2. We adopt that approach for our task and further investigate the impact of token-based document expansion with TILDE on the ranking quality of TILDEv2 in a QBE retrieval setting.

Apart from comparing TILDE and TILDEv2 to the cross-encoder BERT ranker, we also make a comparison to traditional lexical matching models (BM25 and Probabilistic Language models), which have been shown as strong baselines on QBE tasks in prior work~\cite{askari2021combining,rosa2021yes}: 

\begin{enumerate}[RQ2]
\item What is the effectiveness of traditional lexical matching models with varying tokenization strategies in comparison to TILDE and TILDEv2?
\end{enumerate}

To answer \textit{RQ2} we will investigate the effect of using the BERT tokenizer \cite{devlin2019bert} as pre-processing for traditional term-based retrieval models. By doing so, we are aligning the index vocabulary of traditional models with that of TILDE and TILDEv2, which could make our comparison more fair.

We will see in the Section \ref{section:Results} that BM25 shows a competitive ranking quality in comparison to TILDE and TILDEv2 in our QBE benchmark. Because of the similar quality on average, we are interested to see if the relevance signals of TILDE and TILDEv2 are different from that of BM25, to find out if the methods are complementary to each other. To this aim, we will investigate the following research question:

\begin{enumerate}[RQ3]
\item To what extent do TILDE and TILDEv2 encode a different relevance signal from BM25?
\end{enumerate}

To address the question above, as it is described in details in Section \ref{sec:interpolation_describe}, we will analyze the effect of the interpolation of the scores of TILDE and TILDEv2 with BM25.




Since TILDE and TILDEv2 are introduced as re-ranking models, we use four different tasks from the SciDocs evaluation benchmark \cite{cohan2020specter} as a domain-specific QBE benchmark. This benchmark uses scientific paper abstracts as the query and documents. The retrieval setting in these tasks suits as a re-ranking setup because of the number of documents to be ranked for each query. Since that we are working in a domain-specific evaluation setting, we will also address the following research question:
\begin{enumerate}[RQ4]
\item To what extent does a highly tailored domain-specific pre-trained BERT model affect the effectiveness of TILDE and TILDEv2 in comparison to a BERT\textsubscript{base} model?
\end{enumerate}

In summary our main contributions in this work are three-fold:
\begin{itemize}
    \item 
    We show that two recent transformer-based lexical models (TILDE and TILDEv2) are less effective in Query-by-Example retrieval than was expected based on results reported for ad hoc retrieval. This indicates that QBE retrieval is structurally different from other IR settings and requires special attention for methods development; 
    \item  
    We show that the relevance signals of TILDE and TILDEv2 can be complementary to that of BM25 as interpolation of the methods leads to an improvement in ranking effectiveness;
    \item We also investigate interpolations of BM25 with TILDE and TILDEv2 in an ideal setting where the optimal interpolation weight is known a priori, and by doing so, we show that more stratified approaches for the interpolation could result in higher gains from the interpolation of BM25 with TILDE and TILDEv2.
\end{itemize}

In section \ref{sec:Background} we describe the retrieval models used in this work. In section \ref{sec:ExperimentalSettings} we provide details about our methods and experiments and in section \ref{section:Results} we analyze the results and discuss the answers to our research questions. Section \ref{section:discussion} is dedicated to to further analysis of the results, and finally, in Section \ref{section:conclusion} we provide the conclusion.

The code used in this paper is available at: \url{https://github.com/aminvenv/lexica}
\section{Background: Retrieval Models}
\label{sec:Background}
In this section, we briefly introduce the retrieval models that we implement and evaluate in our experiments.
\subsection{Traditional lexical matching models}
\paragraph{BM25}

For BM25 \cite{robertson3m}, we use the implementation by Elasticsearch\footnote{\url{https://github.com/elastic/elasticsearch}} with the parameters \arian{$k=2.75$, and $b=1$}, which was tuned over the validation set.

\paragraph{Probabilistic Language Models}
For language modeling (LM) based retrieval \cite{ponte2017language,hiemstra1998linguistically,berger2017information}, we use the built-in similarity functions of Elasticsearch for the implementation of language model with Jelinek Mercer (JM) smoothing \cite{zhai2004studysmoothing}.

\begin{figure*}
    \centering
    \includegraphics[scale=0.84]{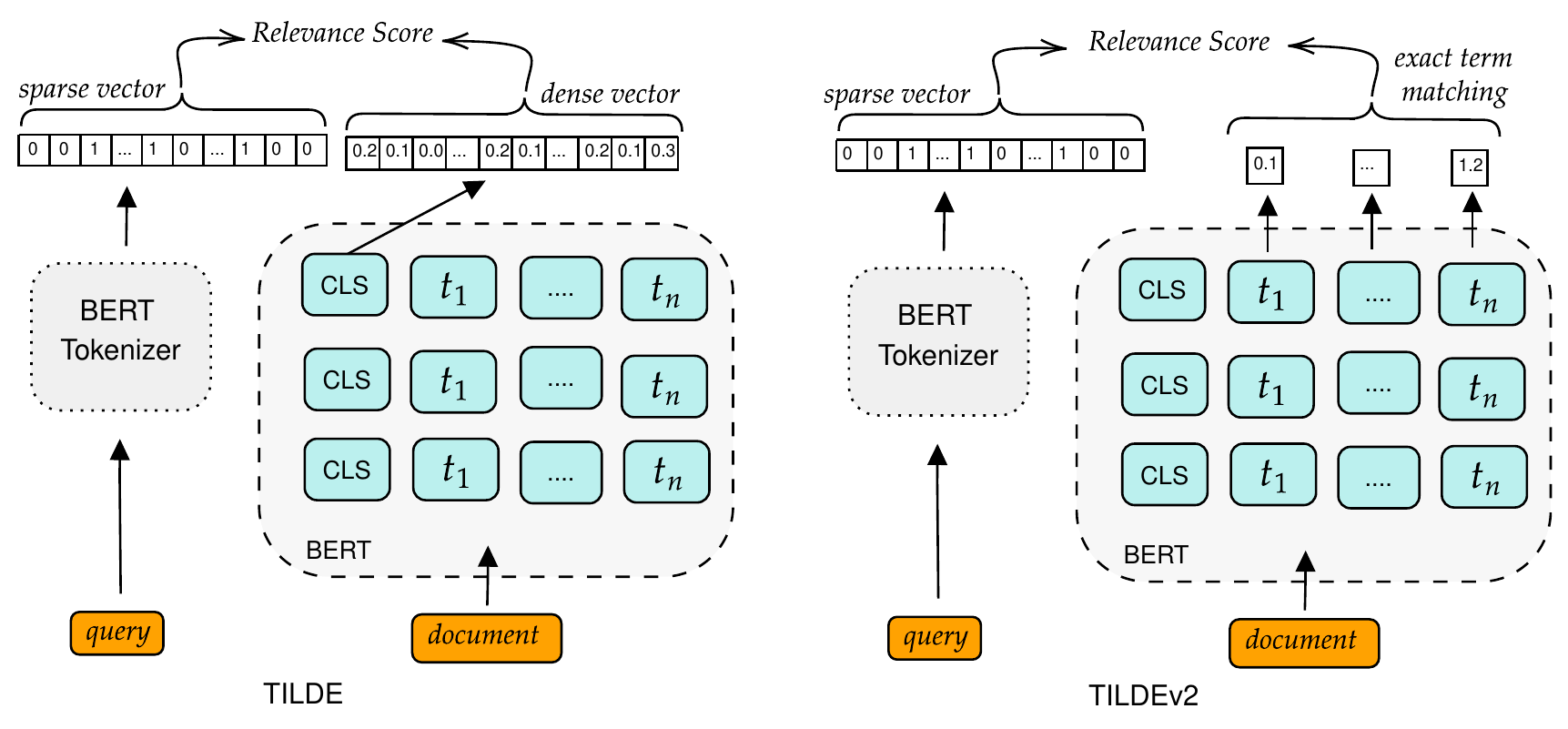}
    \caption{Model architectures. Left: TILDE \cite{zhuang2021tilde}. Right: TILDEv2 \cite{zhuang2021fasttilde}. Both TILDE and TILDEv2 leverage the BERT tokenizer as their query encoder. $t_i$ stands for the $i$th token of the document. The $dense\;vector$ and $sparse\; vector$ have the same length as the BERT vocabulary size.}
    \label{fig:models}
\end{figure*}
\subsection{Term Independent Likelihood
Model: TILDE}
TILDE is a tokenizer-based term-based retrieval model which follows a term independence assumption and formulates the likelihood of a query as follows:
\begin{equation}
    \label{eq:tilde_qlikelihood}
    \text{TILDE-QL}(q|d) = \sum_{i}^{|q|} log(P_{\theta}(q_i|d))
\end{equation}
in which $q$ is the query, and $d$ is the document. As Figure \ref{fig:models} shows, to compute the relevance score, the text of a document $d$  is fed as the input for BERT and the log probability for each token is estimated by using a language modeling head on top of the BERT [CLS] token output. In other words, we are pre-computing the term weights over the complete BERT vocabulary. During both training and inference time, the query text is tokenized by using a BERT tokenizer and the resulting token IDs are used to look up  the corresponding log probability from the likelihood distribution predicted in the output of the language modeling head. It is worth mentioning that the document likelihood can be computed in a similar way by swapping the query and document; however, we only use the query likelihood (Equation \ref{eq:tilde_qlikelihood}) in our experiments.

For TILDE, we use the implementation from the authors' code repository.\footnote{\url{https://github.com/ielab/TILDE}} 
We report results for the TILDE model with different initial checkpoints as the BERT encoder for our fine-tuning procedure. TILDE\textsubscript{BERT} uses bert-base-uncased, TILDE\textsubscript{SciBERT} uses SciBERT, and TILDE\textsubscript{MSMARCO} uses a TILDE which is already fine-tuned on MSMARCO; we use TILDE\textsubscript{MSMARCO} in a zero-shot setting on our data.

\subsection{Lexical Exact Matching: TILDEv2}
TILDE has a drawback in which it expands each document to the size of BERT tokenizer vocabulary. To tackle this problem, the authors proposed TILDEv2. TILDEv2, which builds upon uniCOIL \cite{lin2021few} and TILDE, follows a recent paradigm in contextualized lexical exact matching in which BERT is used to output a scalar importance weight for document tokens \cite{zhuang2021fasttilde,lin2021few}. As it is shown in Figure \ref{fig:models}, in TILDEv2, the token representation is downsized into a scalar weight and the relevance score between a query and a document pair is computed by a sum over the contextualized term weights for all terms appearing in both query and document:
\begin{equation}
    s(q,d) = \sum_{q_i\in{q}}^{} \underaccent{q_i=d_j}{max}(c(q_i) \times v_j^d )
\end{equation}
Here, $q$ and $d$ are the query and the document respectively; $d_j$ is the $j$th token of the document; $v_j^d$ is the term importance weight for the $j$th token of $d$ , and $c(q_i)$ is the count of the $i$-th unique token which is achieved by using the BERT tokenizer as the query encoder. In this equation, $v_j^d$ is computed using the same method as in \citet{lin2021few} in which a $RELU$ function is used on the projection layer to force the model to map the token representations into a positive scalar weight:
\begin{equation}
    v_j^d = ReLU(W_{proj}^{1\times n}BERT(d_j) + b)
\end{equation}
in which $d_j$ is the $j$th token in document $d$ and $b$ is the learnable bias parameter of the projection
layer $W_{proj}$. \citet{lin2021few} show that using a scalar weight as term importance (uniCOIL \cite{lin2021few}) instead of a vector representation (COIL \cite{gao2021coil}) results in a decrease in the effectiveness; however, by using query expansion, uniCOIL can achieve higher effectiveness. Following the method proposed by \citet{zhuang2021fasttilde} for query expansion with TILDE, we will show how TILDEv2 will act when we expand documents with TILDE.
For TILDEv2, we use the implementation from the authors' code repository.\footnote{\url{https://github.com/ielab/TILDE/tree/main/TILDEv2}}

\subsection{Cross-encoder BERT Ranker}
\label{sec:cross-encod}
The state-of-the-art results on SciDocs is reported by  \citet{abolghasemimultitask2022} where they use a multi-task optimized cross-encoder BERT ranker \cite{nogueira2019passage}. The cross-encoder BERT ranker uses the concatenation of query and the document as the input to a BERT encoder. The BERT encoder is then followed by a projection layer $W_{proj}$ on top of its $[CLS]$ token to compute the relevance score:
\begin{equation}
    s(q,d)\;=\; BERT([CLS]\;q\;[SEP]\;d\;[SEP])_{[CLS]}*\;W_{proj}
\end{equation}
In this equation, $q$ and $d$ represent the query and the document respectively and $[CLS]$ as well as $[SEP]$ are special BERT tokens \cite{devlin2019bert}.
\section{Methods and Experimental Settings}
In this section, we provide details and preliminaries about our methods and experimental settings.
\label{sec:ExperimentalSettings}
\begin{figure*}
    \centering
    \includegraphics[scale=1]{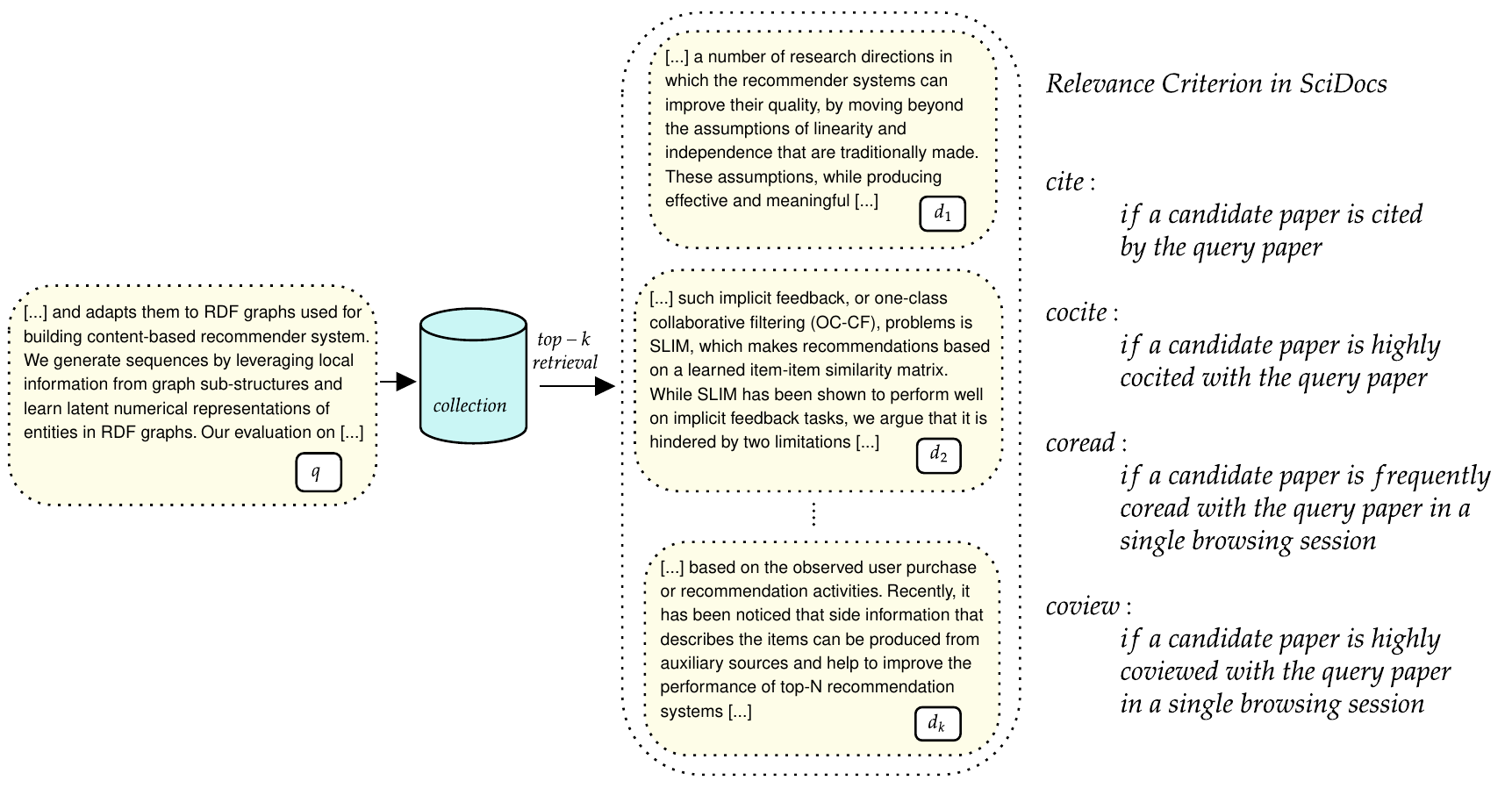}
    \caption{In the Query-by-Example retrieval setting, given a document (in its meaning as a unit of retrieval \cite{lin2021pretrained}) as the query $q$, the goal is to retrieve and rank the top-k relevant documents $\{d_1, \;d_2,\;...\;d_k\}$ out of a collection of documents. We use the four QBE tasks from SciDocs \cite{cohan2020specter} benchmark including $\{cite, cocite, coread, coview\}$, each of which has its own relevance criterion \cite{cohan2020specter}.}
    \label{fig:task_definition}
\end{figure*}
\subsection{Evaluation Benchmark}
We run our experiments on the SciDocs benchmark \cite{cohan2020specter}. This dataset was originally introduced as a benchmark for representation learning tasks. Later, several works including \cite{abolghasemimultitask2022,mysore2021multi} used the tasks of \{co-view, co-read, citation, co-citation\}-prediction from this benchmark as a query-by-example retrieval setting. As Figure \ref{fig:task_definition} depicts, in this setting, given a query document, the goal is to retrieve and rank the most relevant documents out of a collection. The evaluation dataset for each of these four tasks includes approximately 30K total papers from a held-out pool of papers, consisting of 1K query papers and a candidate set of up to 5 positive papers and 25 negative papers \cite{cohan2020specter}.

To make our results comparable, we follow the prior work on SciDocs to prepare the same training data \cite{abolghasemimultitask2022}. To this aim, we take the validation set of each of tasks and use 85\% of them as training and 15\% of them as the validation. Thus, each query in the train set has 5 relevant documents and 25 non-relevant documents. \amin{While TILDE is trained over
relevant query-document pairs \cite{zhuang2021tilde}, TILDEv2 needs triplets in the format of (query, positive document, negative document).} To prepare these triplets we pick two non-relevant documents per relevant document. By doing so, we create 10 triplets out of 30 training samples for each query. It should be noted that following \citet{cohan2020specter} we use a concatenation of abstract and title of the papers as documents.

\subsection{BERT-based Tokenization in Traditional Models}
\label{sec:EXP-Trad}
In order to address \textit{RQ2}, we will examine the effects of transformer-based tokenizers as text pre-processor for traditional retrieval models. Doing so aligns the index vocabulary of
traditional models with that of TILDE and TILDEv2, which in turn
makes our comparison more fair.
Transformers use different tokenization mechanisms e.g. WordPiece  \cite{wu2016googlewpiece}, which result in different query and document representations compared to common word-based tokenization approaches that are sometimes combined with normalization steps such as stemming and lemmatizing. \citet{kamps2020impact} show that using the BERT tokenizer as a pre-processor for BM25 results in a higher efficiency at the cost of a small decrease in effectiveness on the TREC 2020 Deep Learning Track \cite{craswell2020trecdl}. QBE retrieval, however, has the challenge of long queries. In this work, investigate whether the same effect applies to a QBE retrieval setting. To this aim, we use the BERT\textsubscript{base} tokenizer as a pre-processor for LM and BM25. 

In addition, we use the SciBERT tokenizer, which is a domain-specific BERT tokenizer, to find out if a domain-specific tokenizer would have a different effect in comparison to the BERT\textsubscript{base} tokenizer. 
We use three different pre-processing setups in Elasticsearch to compare with our two transformer-based tokenizers:
\begin{itemize}
    \item Elasticsearch Standard Analyzer (SA)
    \item Lowercase token filter, Porter Stemmer, Whitespace tokenizer (STM1)
    \item Lowercase token filter, Porter Stemmer, Standard tokenizer (STM2)
\end{itemize}

In Table \ref{table:traditional}, models corresponding to these setups respectively have SA, STM1, and STM2 as their subscript. BERT-Token and SciBERT-Token as subscripts stand for using BERT and SciBERT tokenizers as the text pre-processors.

\subsection{Interpolation between BM25 and TILDE (TILDEv2) scores}
\label{sec:interpolation_describe}
To answer \textit{RQ3} about the difference between BM25 and TILDE (as well as TILDEv2) in terms of their relevance signals, following \citet{wang2021bertinterpolate}, we evaluate the effect of the interpolation between the relevance scores from BM25 and from the contextualized term-based ranking models TILDE and TILDEv2. 
To this aim the interpolated score is computed as following:
\begin{equation}
    s(q,d) = \alpha * s_{BM25}(q,d) + (1-\alpha) * s_{contextualized}(q,d)
\end{equation}
Here, $s_{BM25}$ stands for the BM25 score for query $q$, and document $d$, and $s_{contextualized}$ refers to the relevance score from TILDE or TILDEv2.
Also, $\alpha$ is the hyperparameter that controls the impact of the scores from BM25 and TILDE (or TILDEv2). Prior to the interpolation both of the relevance scores are normalized using $z$-scaling (subtracting the mean and dividing by the standard deviation). We optimize $\alpha$ on the validation set.

Additionally, to further investigate the impact of interpolation, we do a per-query oracle interpolation in which we assume the best interpolation setting, i.e., optimal $\alpha$, could be predicted per query, and thus we can explore how much effectiveness is reachable by the interpolation of the scores. In the following of the paper, ``oracle interpolation'' refers to this latter interpolation setup and ``non-oracle interpolation'' refers to the vanilla interpolation, i.e., one $\alpha$ for all queries that is optimized on the validation set.

\subsection{Document Expansion with TILDE}
\label{sec:expansion-definition}
The Average token count of SciDocs documents (abstract+title) is 219 and 208 for BERT and SciBERT respectively. Their 90\% token count quantiles are 341 and 385. Comparing these numbers to the maximum input length of BERT models, i.e., 512 tokens, we can see a capacity for the expansion of the documents. To further investigate \textit{RQ1}, following recent works which use document expansion to alleviate the vocabulary mismatch in contextualized term-based retrieval \cite{zhuang2021fasttilde,lin2021few}, we evaluate the impact of retrieval on documents which are expanded at indexing time. 

To this aim, we use TILDE in the same way as the original paper \cite{zhuang2021fasttilde}. TILDE is at an advantage where it is more efficient than doc2query \cite{nogueira2019doc2query}. In this work, using TILDE\textsubscript{SciBERT} of which we found it performs the best compared to other TILDE models (Table \ref{table:scidocs-contextualized}), we generate $m =200$, and $m=300$ expansion terms for TILDEv2\textsubscript{SciBERT}. It is noteworthy that similar to the original paper  \cite{zhuang2021fasttilde} not all expansion terms are added to a document, but only new expansion terms --- that are not yet present in the document --- are added.

\subsection{Domain-Specific BERT in TILDE and TILDEv2}
To answer \textit{RQ1}, and \textit{RQ4}, we will investigate the power that can be brought by domain-specific pre-training to term-based ranking models. To do so, we evaluate the models' ranking quality in three settings: a) using BERT\textsubscript{base} as encoder, b) zero-shot utilization of TILDE and TILDEv2 models which are already fine-tuned on MSMARCO, and c) using a domain-specific pre-trained BERT as their encoder. Specifically, we use SciBERT \cite{beltagy2019scibert} since our evaluation benchmark is from the scientific domain.

\subsection{Implementation Details}
We run our experiments on NVIDIA RTX 3090 GPU machines with 24GB GPU memory. For BERT\textsubscript{base}, and SciBERT we use the pre-trained models available on Huggingface. All BERT-based models are trained for 5 epochs. We use the Adam optimizer \cite{kingma2015adam} with a learning rate of $2\times10^{-5}$ for TILDE, and the AdamW optimizer with a learning rate of $5\times10^{-6}$ for TILDEv2. In addition, we relax the maximum document length to the maximum input length of BERT during indexing. 

    \begin{table*}[t]
        \centering   
        
        \caption{Ranking quality on the four SciDocs benchmark tasks using contextualized term-based ranking and cross-encoder BERT. ``BERT'' and ``SciBERT'' refers to the pre-trained model used as the encoder. ``MSMARCO" indicates the utilization of TILDE or TILDEv2 which are already fine-tuned on MSMARCO. Rows $i$ and $j$ refer to the experiments on expanded documents with $m$ terms using TILDE\textsubscript{SciBERT} as described in section \ref{sec:expansion-definition}. Statistical significance improvements are according to paired t-test (p<0.05) with Bonferroni correction for multiple testing. Rows $a$ and $b$ are included from Table \ref{table:traditional} for ease of comparison.}
        \label{table:scidocs-contextualized}
        
        \scalebox{0.8}{
        \setlength{\tabcolsep}{6pt} 
        \renewcommand{\arraystretch}{1.55} 
        \begin{tabular}{l | c | c | c | c |c| c| c | c  }
            
            \multirow{3}{*}{\textbf{Model}} &
            \multicolumn{2}{c|}{\textbf{Co-view}} 
            &
            \multicolumn{2}{c|}{\textbf{Co-read}} 
            &
            \multicolumn{2}{c|}{\textbf{Co-cite}} &
            \multicolumn{2}{c}{\textbf{Cite}} 
            \\
            & \textbf{MAP} & \textbf{nDCG} &  \textbf{MAP} & \textbf{nDCG }& \textbf{MAP }& \textbf{nDCG}  & \textbf{MAP} & \textbf{nDCG}
            \\ \hline 
            a) BM25\textsubscript{STM2} &
            {80.8\%$^{bcdf-j}$} & {0.9032$^{cdfgj}$} &
            {81.31\%$^{dfg}$} & {0.9112$^{cdg}$} &
            {81.53\%$^{bcdfg}$} & {0.9171$^{cdfg}$} &
            {79.74\%$^{bdg}$} & {0.9085$^{dg}$}
            \\ 
            b) BM25\textsubscript{SciBERT-Token} &
            80.08\%$^{cdfg}$ & 0.8992$^{cdg}$ &
            80.97\%$^{dfg}$ & 0.9105$^{dg}$ &
            80.83\%$^{cdfg}$ & 0.9141$^{cdg}$ &
            79.03\%$^{dg}$ & 0.9051$^{dg}$
            \\ \hline
            c) TILDE\textsubscript{BERT} &
             76.74\%$^{d}$ & 0.8761$^{d}$ &
             80.57\%$^{dg}$ & 0.8983$^{d}$ &
             79.7\%$^{dg}$ & 0.8999$^{d}$ &
             82.15\%$^{abdg}$ & 0.914$^{dg}$
            \\ 
            d) TILDE\textsubscript{MSMARCO} &
            68.22\%$^{}$ & 0.8261$^{}$ &
            66.75\%$^{}$ & 0.8206$^{}$ &
            65.21\%$^{}$ & 0.8145$^{}$ &
            65.29\%$^{}$ & 0.8186$^{}$
            \\ 
            e) TILDE\textsubscript{SciBERT} &
            82.6\%$^{a-df-j}$ & 0.9115$^{bcdf-j}$ &
             85.03\%$^{a-df-j}$ & 0.9256$^{a-df-j}$ &
             86.38\%$^{a-df-j}$ & 0.9375$^{a-df-j}$ &
             87.74\%$^{a-df-j}$  & 0.9431$^{a-dfghj}$ 
            \\ 
            \hline
            f) TILDEv2\textsubscript{BERT} &
             79.17\%$^{cdg}$ & 0.8948$^{cd}$ &
             80.16\%$^{dg}$ & 0.9051$^{dg}$ &
             80.22\%$^{dg}$ & 0.9103$^{dg}$ &
             82.54\%$^{abdg}$ & 0.9230$^{abdg}$
            \\
            g) TILDEv2\textsubscript{MSMARCO} &
            77.84\%$^{cd}$ & 0.8876$^{d}$ &
            78.53\%$^{d}$ & 0.8959$^{d}$ &
            78.17\%$^{d}$ & 0.9006$^{d}$ &
            75.62\%$^{d}$ & 0.8866$^{d}$
            \\ 
            h) TILDEv2\textsubscript{SciBERT} &
            79.59\%$^{cdg}$ & 0.8961$^{cdg}$ &
             80.74\%$^{dg}$ & 0.9080$^{dg}$ &
             80.94\%$^{cdfg}$ & 0.9123$^{dg}$ &
             84.18\%$^{a-dfg}$ & 0.9314$^{a-dfg}$
            \\ 
            \hline
            TILDEv2\textsubscript{SciBERT} 
            & & & & & & & &
             \\ 

            
           i) \quad\quad expansion w/ m=200 &
            80.06\%$^{cdfgj}$ & 0.8985$^{cdg}$ & 
            81.29\%$^{dfgh}$ & 0.9096$^{dg}$ & 
            81.62\%$^{cdfg}$ & 0.9153$^{cdg}$ &   
            86.42\%$^{a-dfghj}$ & 0.9412$^{a-dfghj}$
            \\
           j) \quad \quad expansion w/ m=300 &
            79.38\%$^{cdg}$ & 0.8942$^{cd}$ & 
            81.17\%$^{dfg}$ & 0.9099$^{dg}$ & 
            81.93\%$^{bcdfgh}$ & 0.9165$^{cdg}$ &   
            84.4\%$^{a-dfg}$ & 0.9319$^{a-dfg}$ 
            \\
            \hline
            k) Cross-Encoder\textsubscript{SciBERT} &
            85.2\%$^{a-j}$ & 0.925$^{a-j}$ &
            87.5\%$^{a-j}$ & 0.940$^{a-j}$ &
            89.7\%$^{a-j}$ & 0.955$^{a-j}$ &
            94.0\%$^{a-j}$ & 0.975$^{a-j}$ 
            \\
            l) Cross-Encoder\textsubscript{MTFT-SciBERT} &
            \bf{86.2\%}$^{a-j}$ & \bf{0.930}$^{a-j}$ &
            \bf{87.7\%}$^{a-j}$ & \bf{0.940}$^{a-j}$ &
            \bf{91.0\%}$^{a-j}$ & \bf{0.961}$^{a-j}$ &
            \bf{94.2\%}$^{a-j}$ & \bf{0.976}$^{a-j}$ 

            \\ 
            \hline
        \end{tabular}
    }
    \end{table*} 
\section{Results}
\label{section:Results}

        \begin{table*}[t]
        \centering   
        \caption{Ranking quality of traditional retrieval models on the four SciDocs benchmark tasks with different tokenization approaches. SA, STM1, STM2, BERT-Token, and SciBERT-Token refer to the pre-processing setting as described in section \ref{sec:EXP-Trad}. Statistical significance improvements are according to paired t-test (p<0.05) with Bonferroni correction for multiple testing.}
        \label{table:traditional}
        \setlength{\tabcolsep}{6pt} 
        \renewcommand{\arraystretch}{1.55} 
        \scalebox{0.85}{
        \begin{tabular}{l | c | c | c | c |c| c| c | c }
            
            \multirow{3}{*}{\textbf{Model}} &
            \multicolumn{2}{c|}{\textbf{Co-view}} 
            &
            \multicolumn{2}{c|}{\textbf{Co-read}} 
            &
            \multicolumn{2}{c|}{\textbf{Co-cite}} &
            \multicolumn{2}{c}{\textbf{Cite}} 
            \\ 
            & \textbf{MAP} & \textbf{nDCG} &  \textbf{MAP} & \textbf{nDCG }& \textbf{MAP }& \textbf{nDCG}  & \textbf{MAP} & \textbf{nDCG}
            \\ \hline

            a) LM\textsubscript{SA} &
                74.78\%$^{}$ & 0.8724$^{}$ &
                74.32\%$^{b}$ & 0.8750$^{}$ &
                74.64\%$^{}$ & 0.8812$^{}$ &
                71.30\%$^{}$ & 0.8653$^{}$
            \\  
            b) LM\textsubscript{STM1} &
            74.82\%$^{}$ & 0.8737$^{}$ &
            73.51\%$^{}$ & 0.8694$^{}$ &
            74.60\%$^{}$ & 0.8810$^{}$ &
            70.98\%$^{}$ & 0.8636$^{}$
            \\ 
            c) LM\textsubscript{STM2} &
            75.74\%$^{abde}$ & 0.8786$^{abe}$ &
            74.90\%$^{ab}$ & 0.8771$^{b}$ &
            75.80\%$^{abde}$ & 0.8873$^{ab}$ &
            72.15\%$^{abe}$ & 0.8696$^{b}$
            \\  
            d) LM\textsubscript{BERT-Token} &
            74.9\%$^{}$ & 0.8734$^{}$ &
            74.76\%$^{b}$ & 0.8778$^{b}$ &
            74.95\%$^{}$ & 0.8829$^{}$ &
            72.04\%$^{abe}$ & 0.8694$^{}$
            \\ 
            e) LM\textsubscript{SciBERT-Token} &
            74.74\%$^{}$ & 0.8717$^{}$ &
            74.69\%$^{b}$ & 0.8771$^{b}$ &
            74.81\%$^{}$ & 0.8827$^{}$ &
            71.46\%$^{}$ & 0.8666$^{}$
            \\ \hline
            f) BM25\textsubscript{SA} &
                77.86\%$^{a-e}$ & 0.8876$^{a-e}$ &
                78.03\%$^{a-e}$ & 0.8949$^{a-e}$ &
                77.95\%$^{a-e}$ & 0.8994$^{a-e}$ &
                76.12\%$^{a-e}$ & 0.8892$^{a-e}$
            \\ 
            g) BM25\textsubscript{STM1} &
            80.21\%$^{a-f}$ & 0.9002$^{a-f}$ &
            80.52\%$^{a-f}$ & 0.9074$^{a-f}$ &
            80.85\%$^{a-f}$ & 0.9137$^{a-f}$ &
            79.03\%$^{a-f}$ & 0.9048$^{a-f}$
            \\
            h) BM25\textsubscript{STM2} &
            \textbf{80.8\%}$^{a-gij}$ & \textbf{0.9032}$^{a-gi}$ &
            \textbf{81.31\%}$^{a-gi}$ & \textbf{0.9112}$^{a-g}$ &
            \textbf{81.53\%}$^{a-gij}$ & \textbf{0.9171}$^{a-g}$ &
            \textbf{79.74\%}$^{a-gij}$ & \textbf{0.9085}$^{a-g}$
            \\
            i) BM25\textsubscript{BERT-Token} &
            79.76\%$^{a-f}$ & 0.8974$^{a-f}$ &
            80.61\%$^{a-f}$ & 0.9088$^{a-f}$ &
            80.5\%$^{a-f}$ & 0.9125$^{a-f}$ &
            79.19$^{a-f}$ \% & 0.9057$^{a-f}$
            \\ 
            j) BM25\textsubscript{SciBERT-Token} &
            80.08\%$^{a-f}$ & 0.8992$^{a-f}$ &
            80.97\%$^{a-gi}$ & 0.9105$^{a-f}$ &
            80.83\%$^{a-fi}$ & 0.9141$^{a-f}$ &
            79.03\%$^{a-f}$ & 0.9051$^{a-f}$
            \\ 
\hline
    \end{tabular}
    }
    \end{table*}  
 
 \begin{table*}[t]
        \centering   
        
        \caption{Results for non-oracle interpolation (the interpolation parameter $\alpha$ is optimized on the validation set) between BM25$_{STM2}$, TILDE$_{SciBERT}$, and TILDEv2$_{SciBERT}$. Statistical significance improvements are according to paired t-test (p<0.05) with Bonferroni correction for multiple testing. Rows $a$, $b$, and $c$ are included from Table \ref{table:scidocs-contextualized} for ease of comparison.}
        \label{table:non-oracle}
        
        \scalebox{0.8}{
        \setlength{\tabcolsep}{6pt} 
        \renewcommand{\arraystretch}{1.55} 
        \begin{tabular}{l | c | c | c | c |c| c| c | c  }
            
            \multirow{3}{*}{\textbf{Model}} &
            \multicolumn{2}{c|}{\textbf{Co-view}} 
            &
            \multicolumn{2}{c|}{\textbf{Co-read}} 
            &
            \multicolumn{2}{c|}{\textbf{Co-cite}} &
            \multicolumn{2}{c}{\textbf{Cite}} 
            \\
            & \textbf{MAP} & \textbf{nDCG} &  \textbf{MAP} & \textbf{nDCG }& \textbf{MAP }& \textbf{nDCG}  & \textbf{MAP} & \textbf{nDCG}
            \\ \hline 
            a) BM25\textsubscript{STM2} &
            {80.8\%$^{c}$} & {0.9032$^{c}$} &
            {81.31\%$^{}$} & {0.9112$^{}$} &
            {81.53\%$^{}$} & {0.9171$^{}$} &
            {79.74\%$^{}$} & {0.9085$^{}$}
            \\ 
            \hline
            b) TILDE\textsubscript{SciBERT} &
            82.6\%$^{ac}$ & 0.9115$^{c}$ &
             85.03\%$^{ace}$ & 0.9256$^{ac}$ &
             86.38\%$^{ace}$ & 0.9375$^{ace}$ &
             87.74\%$^{ace}$  & 0.9431$^{ace}$ 
            \\ 
            \hline
           
            c) TILDEv2\textsubscript{SciBERT} &
            79.59\%$^{}$ & 0.8961$^{}$ &
             80.74\%$^{}$ & 0.9080$^{}$ &
             80.94\%$^{}$ & 0.9123$^{}$ &
             84.18\%$^{a}$ & 0.9314$^{a}$
            \\ 
            \hline
            d) BM25\textsubscript{STM2} + TILDE\textsubscript{SciBERT} &
            {\textbf{85.29\%}$^{abce}$} & {\textbf{0.9214}$^{abce}$} &
            {\textbf{86.52\%}$^{abce}$} & {\textbf{0.9395}$^{abce}$} &
            {\textbf{88.32\%}$^{abce}$} & {\textbf{0.9494}$^{abce}$} &
            {\textbf{88.46\%}$^{abce}$} & {\textbf{0.9496}$^{abce}$}
            \\
            e) BM25\textsubscript{STM2} + TILDEv2\textsubscript{SciBERT} &
            {81.56\%$^{ac}$} & {0.9032$^{c}$} &
            {82.63\%$^{ac}$} & {0.9183$^{ac}$} &
            {83.06\%$^{ac}$} & {0.9242$^{ac}$} &
            {84.18\%$^{a}$} & {0.9318$^{a}$}
            \\
            \hline

        \end{tabular}
    }
    \end{table*} 
    
\begin{figure*}
    \centering
    \includegraphics[scale=0.45]{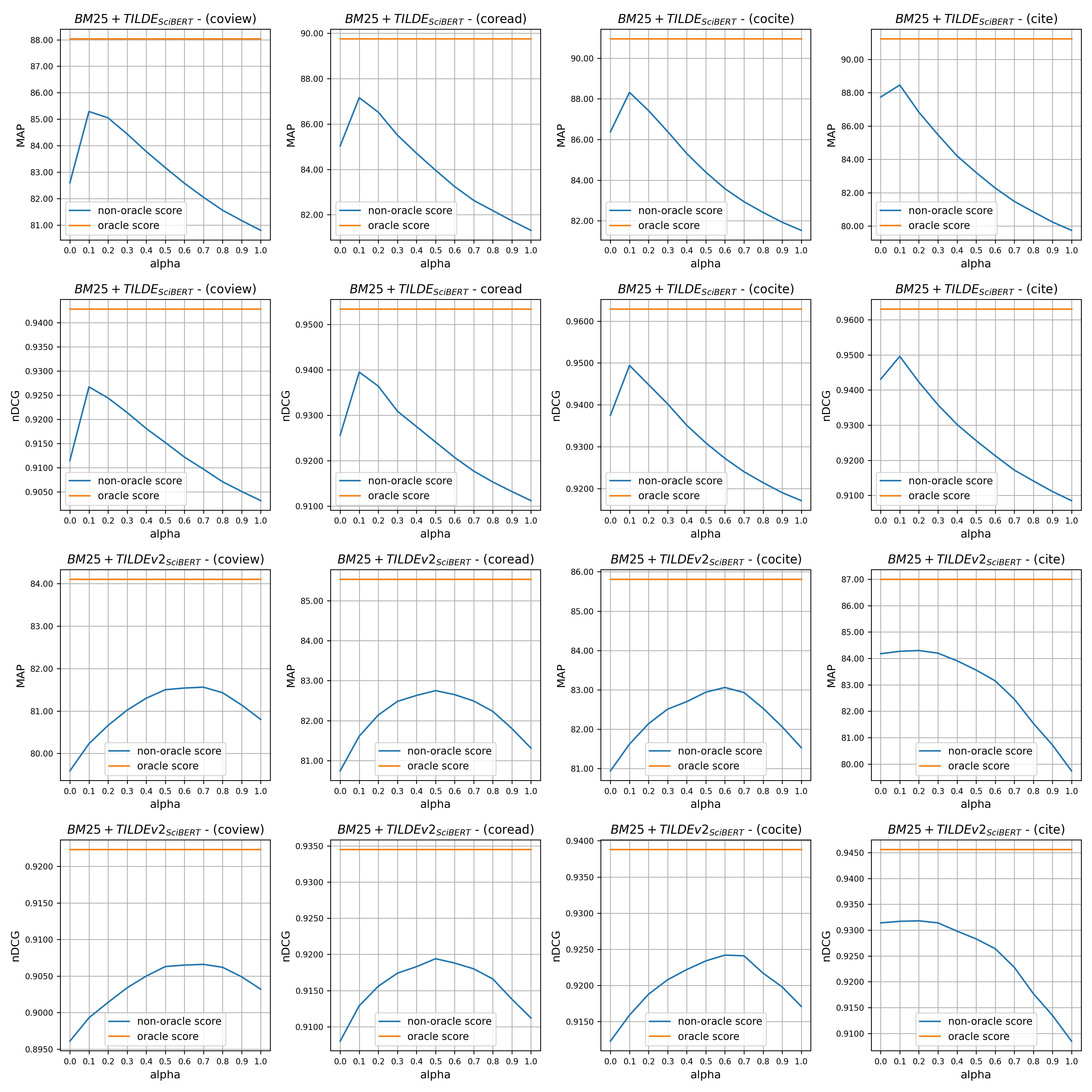}
    \caption{Results for TILDE and TILDEv2 with varying values of interpolation parameter $\alpha$. The lines in blue and orange represent the effectiveness based on the non-oracle and oracle interpolations respectively. $\alpha=0.0$ represents the TILDE- or TILDEv2-only setting; $\alpha=1.0$ represents the BM25-only setting.}
    \label{fig:interpolation}
\end{figure*}

        
        
    
\paragraph{\textbf{RQ1}} \textit{How effective are TILDE and TILDEv2 in query-by-example retrieval? and 
\textbf{RQ4} To what extent does a highly tailored domain-specific pre-trained BERT model affect the effectiveness of TILDE and TILDEv2 in comparison to when we use a BERT\textsubscript{base} model?}

As Table \ref{table:scidocs-contextualized} shows, TILDE and TILDEv2 are less effective than a cross-encoder BERT ranker in QBE retrieval despite having longer queries. This could be due to the fact that the cross-encoder BERT ranker applies all-to-all attention across tokens in both the query and the document \cite{lin2021pretrained} and thus, query terms and document terms are highly contextualized for the estimation of the relevance score. In addition, we see that TILDEv2\textsubscript{BERT} outperforms TILDE\textsubscript{BERT} despite TILDEv2 being highly prune to the vocabulary mismatch problem. One hypothesis for this observation could be that in a domain-specific retrieval setup like ours, TILDEv2 with the BERT\textsubscript{base} encoder predicts more effective document term weights than the term weights predicted for all tokens in the BERT vocabulary by TILDE with the BERT\textsubscript{base} encoder.

In addition, using SciBERT as our domain-specific pre-trained BERT model unsurprisingly improves the ranking quality of both TILDE and TILDEv2; however, this improvement is higher between TILDE\textsubscript{BERT} and TILDE\textsubscript{SciBERT} than between TILDEv2\textsubscript{BERT} and TILDEv2\textsubscript{SciBERT} to an extent where TILDE\textsubscript{SciBERT} even outperforms both TILDEv2\textsubscript{BERT} and TILDEv2\textsubscript{SciBERT}. This observation could be due to the fact that the vocabulary mismatch problem caused by exact matching limits the TILDEv2 ranking quality, even if we use a highly tailored domain-specific BERT as its encoder. In this respect, we investigate the impact of token-based query expansion (see section \ref{sec:expansion-definition}) with TILDE on the ranking quality of TILDEv2 in our QBE retrieval setting. Lines $i$, and $j$ in Table \ref{table:scidocs-contextualized} are the ranking results on the documents that are expanded using TILDE with the method introduced by \citet{zhuang2021fasttilde}. Here, we are interested to find out if using document expansion is able to compensate for the gap in the ranking quality between TILDE\textsubscript{SciBERT}, and TILDEv2\textsubscript{SciBERT}.

As shown in Table \ref{table:scidocs-contextualized}, TILDEv2\textsubscript{SciBERT} with $m=\{200,300\}$ expansion terms, is still less effective than TILDE\textsubscript{SciBERT}.  Furthermore, the table shows that the ranking quality of BM25$_{STM2}$ on the original documents (line $a$) is still comparable with the ranking quality of TILDEv2\textsubscript{SciBERT} on the expanded documents (lines $i$ and $j$). It is noteworthy that to make sure we are expanding the documents with enough tokens we investigate the average number of tokens added to the documents by the expansion with TILDE\textsubscript{SciBERT}. By doing so, we find that for $m=200$, and $m=300$, approximately 49 and 128 new tokens are appended to the documents on average. Additionally, we find that using $m=100$ results in roughly 2.6 new tokens on average. These numbers beside the statistics of the tokens in SciDocs benchmark, provided in Section \ref{sec:expansion-definition}, indicate that $m$ should be tuned in order to take advantage from the document expansion with TILDE in QBE retrieval setting. \amin{ Finally, we see that the zero-shot utilization of TILDE\textsubscript{MSMARCO} and TILDEv2\textsubscript{MSMARCO} does not show superior performance over the fine-tuned TILDE and TILDEv2 with both BERT and SciBERT encoders. It should be noted that taking models which are already fine-tuned on general domain (like TILDE\textsubscript{MSMARCO} and TILDEv2\textsubscript{MSMARCO}) and further fine-tuning them on the task domain is a typical approach which could result in improvement in their ranking quality; however, we leave this item as a direction to be explored in future work}.


\paragraph{\textbf{RQ2}} \textit{What is the effectiveness of traditional lexical matching models with varying tokenization strategies in comparison to TILDE and TILDEv2}

Table \ref{table:traditional} shows that leveraging BERT and SciBERT tokenizers results in competitive ranking quality in both probabilistic language model based retrieval and BM25 in comparison to the three traditional pre-processing setups introduced in section \ref{sec:EXP-Trad}.

Moreover, as the results of Table \ref{table:scidocs-contextualized} shows, the ranking quality of BM25$_{STM2}$ not only outperforms LM and BM25 with different traditional and BERT-based pre-processing
approaches, but also it could even outperform TILDE$_{BERT}$, and TILDEv2$_{BERT}$ in most of the tasks. In fact, we do not see a large gap between BM25 compared to TILDEv2 as was shown for retrieval based on short queries in the experiments on MSMARCO and TREC DL Track benchmarks \cite{zhuang2021fasttilde}. This finding is important as (1) it sheds light on the challenges of retrieval settings different from the common evaluation benchmarks including MSMARCO and the TREC DL Track; (2) raises the question how effective other contextualized term-based ranking models would be in those settings.

\begin{table*}[t]
        \centering
        \caption{Results for oracle interpolation (optimal $\alpha$ per query) between BM25$_{STM2}$, TILDE$_{SciBERT}$, and TILDEv2$_{SciBERT}$. Statistical significance with paired t-test (p<0.05) is reported only with respect to non-interpolated scores ($\alpha$=0) of these three models in Table \ref{table:scidocs-contextualized} (row a, e and h). $\alpha$\textsubscript{average} represents the mean of the optimal $\alpha$ values picked per query. The number of queries with optimal $\alpha$=0 stands for the number of queries for which the interpolation does not improve their effectiveness compared to TILDE or TILDEv2 only.}
        \label{table:oracle}
        \setlength{\tabcolsep}{5pt} 
        \renewcommand{\arraystretch}{1.25} 
        \scalebox{0.8}{
        \begin{tabular}{l | c | c | c | c |c| c| c | c }
            
            \multirow{3}{*}{\textbf{Model}} &
            \multicolumn{2}{c|}{\textbf{Co-view}} 
            &
            \multicolumn{2}{c|}{\textbf{Co-read}} 
            &
            \multicolumn{2}{c|}{\textbf{Co-cite}} &
            \multicolumn{2}{c}{\textbf{Cite}} 
            \\ 
            & \textbf{MAP} & \textbf{nDCG} &  \textbf{MAP} & \textbf{nDCG }& \textbf{MAP }& \textbf{nDCG}  & \textbf{MAP} & \textbf{nDCG}
            \\ \hline

            BM25 $+_{oracle}$ TILDE$_{SciBERT}$ &
            & & & & & & &
            \\
            \quad ranking quality &
               88.04\%$^{ae}$ & 0.9428$^{ae}$ &
               89.76\%$^{ae}$ & 0.9534$^{ae}$ &
               90.96\%$^{ae}$ & 0.9629$^{ae}$ &
               91.24\%$^{ae}$ & 0.9631$^{ae}$
                \\
            \quad $\alpha$\textsubscript{average} &
             0.1265 & 0.1294 &
             0.1048 & 0.1053 &
             0.1044 & 0.1048 & 
             0.0839 & 0.0857
                \\
            \quad \#queries with optimal $\alpha$=0 &
            537 & 533 &
            563 & 557 & 
            550 & 552 &
            624 & 619
               \\
           \quad \#queries with optimal $\alpha$=1 &
           19 & 21 & 
           6 & 5 &
           14 & 13 & 
           4 & 4
            \\
            \quad  IQR of the optimal $\alpha$ over queries &
            0.1 & 0.2 &
            0.1 & 0.1 & 
            0.1 & 0.1 & 
            0.1 & 0.1
            \\  \hline
            BM25 $+_{oracle}$  TILDEv2$_{SciBERT}$ &
            & & & & & & &
            \\
            \quad ranking quality &
            84.10\%$^{ah}$ & 0.9223$^{ah}$ &
            85.54 \%$^{ah}$ & 0.9345$^{ah}$ &
            85.81 \%$^{ah}$ & 0.9388$^{ah}$ &
            87.00 \%$^{ah}$ & 0.9456$^{ah}$ 
            \\
            \quad $\alpha$\textsubscript{average} &
            0.3169 & 0.3205 & 
            0.3040 & 0.3048 & 
            0.3337 & 0.3339 & 
            0.2073 & 0.2083
            \\
            \quad \#queries with optimal $\alpha$=0 &
            467 & 463 & 
            446 & 447 &
            419 & 420 &
            575 & 573
            \\ 
            \quad \#queries with optimal $\alpha$=1 &
            84 & 84 & 
            71 & 72 &
            92 & 94 & 
            33 & 33
            \\
            \quad IQR of the optimal $\alpha$ over queries &
            0.7 & 0.7 & 
            0.6 & 0.6 & 
            0.6 & 0.6 & 
            0.4 & 0.4
            \\
\hline
    \end{tabular}
    }
    \end{table*}  
    
\paragraph{\textbf{RQ3}}\textit{To what extent do TILDE and TILDEv2 encode a different relevance signal from BM25?}

The blue lines in Figure \ref{fig:interpolation} show the ranking quality for TILDE\textsubscript{SciBERT} and TILDEv2\textsubscript{SciBERT} when their scores are interpolated with the BM25 score over varying values of interpolation parameter $\alpha$ with the step of 0.1. Besides, Table \ref{table:non-oracle} shows the ranking quality for the interpolations with the $\alpha$ that is tuned over the validation set. We can see that an optimal interpolation between the scores from BM25 and the contextualized term-based ranking models TILDE and TILDEv2 could provide significant improvements for almost all tasks over the individual rankers participating in the interpolation. The only exceptions are in the \textit{co-view}, and \textit{cite} tasks. To be specific, there is no improvement over BM25 in the nDCG metric in the \textit{co-view} (line e vs. line a in Table \ref{table:non-oracle}). Besides, in the \textit{cite} task the improvement over TILDEv2 (line e vs. line c in Table \ref{table:non-oracle}) is not significant for the nDCG metric, and there is no improvement for the MAP metric.
Nevertheless, the improvements obtained by the interpolation for almost all tasks and metrics indicates that TILDE and TILDEv2 are capturing different relevance signals compared to BM25.

To further investigate the impact of the score interpolation with BM25 scores, we perform an oracle interpolation in which we assume the optimal interpolation hyperparameter $\alpha$ is known for each individual query. This query-specific optimal value is selected over varying values of $\alpha$ with the step of 0.1. Table \ref{table:oracle} as well as orange lines in Figure \ref{fig:interpolation} show the results for the oracle interpolation. We can see that the oracle interpolation would result in a substantial improvement for both TILDE and TILDEv2. 

Moreover, we can see in Table \ref{table:oracle} that there is a subset of queries for which the BM25 ranking alone is better than the interpolation (queries with optimal $\alpha$=1). This number is lower for the interpolation with TILDE than for the interpolation with TILDEv2. One hypothesis for this observation could be that the interpolation with TILDE is likely to be more helpful for BM25 since TILDE could bring more contextualization power for BM25 as it incorporates the term importance for all tokens in the query. In other words, since TILDEv2 pre-computes term weights only for the tokens of the document (whereas TILDE pre-computes the term importance weight for all the tokens in the BERT vocabulary per document), due to the chance of vocabulary mismatch in TILDEv2, it could incorporate less query-dependent contextualization than TILDE.

In addition, we see that the margin between the oracle interpolation results and both non-interpolated scores as well as non-oracle interpolation scores (Table \ref{table:non-oracle}) is substantial, which demonstrates that more complex aggregation methods could benefit more from the relevance signals from TILDE, TILDEv2 and BM25.

\section{Discussion}
\label{section:discussion}
In this section, we further analyze the interpolation between BM25 and TILDE (TILDEv2) in terms of the interpolation effectiveness and  the interpolation weight $\alpha$.
\subsection{Interpolation effectiveness}
The first two rows on the top of Figure \ref{fig:interpolation} correspond to the interpolation between TILDE and BM25 and the two rows in the bottom correspond to the interpolation between TILDEv2 and BM25. Comparing the nDCG and MAP plots for the interpolation between TILDE and BM25, we can see that for this combination, $\alpha$=0.1 shows the highest ranking quality for both nDCG and MAP metrics in all tasks. Thus, a high weight for TILDE with a small weight for BM25 gives the highest effectiveness for this combination. This observation could mean that while TILDE, as contextualized transformer-based model, is able to outperform BM25 as an exact matching model, it could still benefit from the strong lexical relevance scores from BM25. 

On the other hand, for the combination of TILDEv2 and BM25 we see that the highest ranking quality is obtained with $\alpha$ $\in$ \{0.3, 0.4, 0.5, 0.6, 0.7, 0.8\} depending on the task. The exceptions are in the \textit{cite}, and \textit{coview} tasks as described in the answer to \textit{RQ3} in Section \ref{section:Results}. Thus, in the combination of TILDEv2 and BM25, an equal or slightly higher weight for BM25 relative to TILDEvs gives the optimal results. A hypothesis for this observation could be that while both BM25 and TILDEv2 are performing based on exact matching, the term weights from TILDEv2, which are predicted through contextualization of the document terms, are not always more effective than the term scores from BM25; however, they can act as a complement for each other and thus their interpolation could benefit from both.

\subsection{Interpolation weight}
To further analyze the interpolation weight $\alpha$, we consider the two aforementioned settings of oracle interpolation and non-oracle interpolation.
\paragraph{Non-oracle interpolation}
We can see in Figure~\ref{fig:interpolation} (blue lines) that for the effective interpolations, i.e, the interpolations that result in higher effectiveness than each individual ranker included in the interpolation, the interpolation weight $\alpha$ in the combination of BM25 and TILDEv2 has a wider range than in the combination of BM25 and TILDE. This indicates that in this experimental setting the interpolation of BM25 and TILDEv2 could be achieved by a broader range of $\alpha$ values and is therefore more robust to the choice of interpolation weight than for BM25 and TILDE.

\paragraph{Oracle interpolation}
As a measure of the statistical dispersion, we report the inter-quartile range (IQR) for the oracle interpolation weight $\alpha$ which is shown in Table \ref{table:oracle}. Taking the range of $\alpha$ [0.0, 1.0] into account, we can see that we have low inter-quartile range (IQR) for the optimal values of $\alpha$ per query in the interpolation with \textit{TILDE} (top part of the table). On the other hand, the IQR for the optimal values of $\alpha$ per query for the interpolation with \textit{TILDEv2} are much higher (bottom part of the table), which indicates that the optimal interpolation setting for the queries are more varied. This observation could give some sense of robustness against query variation for TILDE in comparison to TILDEv2 in this experimental setting. In other words, a query-dependent approach for optimizing $\alpha$ would be more robust against query variation for TILDE than for TILDEv2.
\section{Conclusion}
\label{section:conclusion}
In this paper we investigated the generalizability of two contextualized term-based ranking models TILDE and TILDEv2 for a QBE retrieval setting. In QBE, the queries are much longer than in ad-hoc retrieval, and efficient query processing is essential. We were specifically interested to see to what extent the relative performance of contextualized term-based ranking models in comparison to both traditional term-based models and the effective cross-encoder BERT ranker is generalizable to a QBE retrieval setting. 

Our results show that similar to the original papers \cite{zhuang2021tilde,zhuang2021fasttilde}, TILDE and TILDEv2 are less effective than a
cross-encoder BERT ranker in QBE retrieval despite the context of longer queries. On the other hand, in the original papers, TILDE and TILDEv2 have shown superior ranking quality in comparison to BM25 as a traditional term-based retrieval model. We investigated if the same pattern exists in a query-by-example retrieval setting and our results show that BM25 has a competitive ranking quality compared to TILDE and TILDEv2. In fact, not only is it competitive, but also in some cases it could outperform TILDE and TILDEv2. 

This finding is important as (1) it sheds light on the challenges of retrieval settings different from the common evaluation benchmarks including MSMARCO and the TREC DL Track; (2) raises the question how effective other contextualized term-based ranking models would be in those settings. Our results indicate that QBE retrieval is structurally different from other IR settings and requires special attention for methods development. 

Furthermore, we investigated the impact of the interpolation between BM25 and TILDE as well as TILDEv2. By doing so, we find that a linear interpolation between the score of TILDE (TILDEv2) with that of BM25 leads to an improvement in the ranking effectiveness. This shows that the relevance signals from contextualized ranking models TILDE and TILDEv2 are complementary to the relevance signals from BM25. Additionally, through an analysis on the oracle interpolation between BM25 and TILDE (TILDEv2), we show that more stratified approaches could benefit more from the interpolation between the scores from these models. 

\section{Acknowledgments}
This work is funded by the DoSSIER project under European Union’s Horizon 2020 research and innovation program, Marie Skłodowska-Curie grant agreement No. 860721.

\bibliographystyle{ACM-Reference-Format}
\balance 
\bibliography{sample-base}

\appendix

\end{document}